\documentclass[11pt]{article}
\pdfoutput=1
\usepackage{jheppub}
\usepackage{amssymb,amsfonts}
\usepackage{mathrsfs}
\let\savenumberline\numberline
\def\numberline#1{\savenumberline{#1.}}
\makeatletter
\renewcommand{\@seccntformat}[1]{\csname the#1\endcsname.\,\,}
\makeatother

\newcommand{\CB}{{\cal B}}

\newcommand{\CH}{{\cal H}}

\newcommand{\CL}{{\cal L}}
\newcommand{\CM}{{\cal M}}
\newcommand{\CN}{{\cal N}}

\newcommand{\MR}{{\mathbb R}}

\renewcommand{\hat}[1]{\widehat{#1}}
\newcommand{\be}{\begin{equation}}
\newcommand{\ee}{\end{equation}}
\newcommand{\bea}{\begin{eqnarray}}
\newcommand{\eea}{\end{eqnarray}}

\newcommand\secref[1]{{\S\ref{#1}}}

\newcommand{\su}{{\mathfrak{su}(2)}}
\makeatletter
\def\@fpheader{\relax}
\makeatother
\usepackage{graphicx}
\usepackage{latexsym}

\title{\ \vspace{1.5in} \\ \hbox{A Tropical Look at Coisotropic Branes and Quantization}}
\author{Emil Albrychiewicz, Andr\'{e}s Franco Valiente, and Vi Hong}
\affiliation{\medskip
Berkeley Center for Theoretical Physics and Department of Physics\\
University of California, Berkeley, CA, 94720-7300, USA\medskip\\
Theoretical Physics Group, Lawrence Berkeley National Laboratory\\
Berkeley, CA 94720-8162, USA}
\emailAdd{ealbrych@berkeley.edu}
\emailAdd{andresfranco@berkeley.edu}
\emailAdd{vihong14@berkeley.edu}
\abstract{We continue our investigation of tropical branes by exploring the tropicalization of topological sigma models with boundaries. We show that the tropical limit naturally decomposes conventional A-branes into two distinct classes: tropical Lagrangian branes and tropical coisotropic branes. By carefully analyzing the modified boundary conditions emerging from the tropological sigma models, we construct these tropical branes explicitly and demonstrate their utility in an example where we quantize a symplectic manifold through the use of a tropical version of brane quantization.}

\begin{document}
\maketitle

\section{Introduction}
Recently, in \cite{trsm}, it has been shown that one can construct topological sigma models that are built upon geometries that are associated with degenerated punctured Riemann surfaces known as tropical geometries. 
The construction of these tropical path integrals are based on a crucial observation that one is not forced to represent a tropical geometry as a real algebraic variety but can instead treat the tropical geometry as a foliated complex geometry. Consequently, the tropical quantum field theories that arise are of the cohomological type with an additional non-relativistic Lifshitz-like symmetry \cite{mqc, qcym}. The upshot of this construction is that we are able to use standard differential-geometric methods in order to investigate whatever questions of interest arise.  

In the last few decades, it has been shown that it is interesting to explore cohomological field theories as a tool in understanding quantum field theory as a whole and as well as an idea generator for formulating unexpected deep connections in mathematics. A natural follow up question to \cite{trsm} is whether one is able to construct interesting invariants through the use of tropical path integrals which do not necessarily arise from a singular limit of a standard topological sigma model but instead come from different quantum field theories. One of possible paths is to try to formulate a geometric differential operator whose index can be shown to be a quantity that is stable under deformations yet interesting enough to carry some geometric and/or topological information about the underlying geometry. We will take this approach in an upcoming paper \cite{WIP1} through use of supersymmetric tropical worldlines and their associated index theorems. 

In this paper, instead, we will take a more pedestrian approach and study the Hilbert spaces that naturally arise upon the quantization of a symplectic manifold.  In particular,  we study the behavior of Hilbert space invariants, such as the dimension, under tropicalization through the relationship that arises between the recently proposed tropological sigma A model \cite{trsm} and  the analytic continuation of the path integral of quantum mechanics \cite{Witten:2010zr}. The analytic continuation uses a Morse-theoretic construction that leads to a construction of a path integral which localizes on the space of psuedoholomorphic maps with specified boundary conditions, naturally leading to topological A branes. We will find that upon tropicalization, we obtain modified boundary conditions giving defining equations for tropical topological A branes, which we will call \textit{tropological} A branes. As in the non-tropical case, we will find that these branes come in two distinct types. The first type is tropical Lagrangian A-branes \cite{Witten:1992fb} and the second will be the tropical coisotropic A-branes \cite{Kapustin:2001ij}. Furthermore, the analytic continuation of the path integral leads to an alternative quantization procedure of symplectic manifolds by embedding it into a string-theoretic setup. One can then study stable invariants such as the dimension of the Hilbert space and index theorems through simpler calculations that arise via various geometric engineering setups and dualities in string theory; this is known as \textit{brane quantization} \cite{Gukov:2008ve}. 

The motivations for this paper are not strictly mathematical but are meant to probe ingredients needed to construct a worldsheet formulation of a non-equilibrium string perturbation theory. In \cite{neq, ssk, keq}, it was stated that the foundational premise of string theory is predicated on the assumption of a stable, static, and eternal vacuum. However, it is well-established that our universe exists in a state far from thermal equilibrium. Consequently, there is a pressing need for a systematic reformulation of string theory that facilitates the exploration of non-equilibrium scenarios. 

Non-equilibrium field theory can generically be formulated in terms of a closed time contour known as the Schwinger Keldysh contour \cite{schwinger1961brownian, keldysh1964diagram}, in which the system is first evolved forward in time and then subsequently backwards in time. The question of how covariant string perturbation theory can be defined on a Schwinger Keldysh time contour was first explored in \cite{neq,ssk,keq} with the use of expected holographic dualities of gauge theories in the large $N$ expansion. It was found that on the Schwinger Keldysh time contour, the string worldsheet $\Sigma$ now carries a natural triple decomposition,
$$
\Sigma=\Sigma^{+} \cup \Sigma^{\wedge} \cup \Sigma^{-}
$$
where $\Sigma^{+}$ corresponds to the part of the worldsheet on the forward branch of the time contour, $\Sigma^{-}$ corresponds to the backward part of the time contour, and the ``wedge region" $\Sigma^{\wedge}$ is the portion of the worldsheet that connects them, and corresponds to the turn-around late time where the forward and backward branch of the time contour meet. Surprisingly, the wedge region is itself a topologically two-dimensional region with arbitrarily complicated topology and its own genus expansion. Geometrically, however, it is expected to be highly anisotropic on the worldsheet, in such a way that the worldsheet develops a nonrelativistic foliation structure where the leaves of foliation connect points on the boundary with the $\Sigma^{+}$ region to a point on the boundary with the $\Sigma^{-}$ region. The immediate question one might ask is: ``How can this wedge region be explicitly constructed in a manner suitable for worldsheet calculations?"  

To make progress in this direction, it is necessary to provide a prescription for addressing the fact that string perturbation theory would need to be defined on analytically continued worldsheets, which are permitted to evolve both forward and backward in time. It is well known that dealing with a covariant and strictly Lorentzian string perturbation theory can lead to uncontrollable singularities upon integrating over all worldsheet geometries and hence it was argued in \cite{Witten:2013pra}, that one can minimally modify the string perturbation theory by allowing the worldsheets to be generically described by a Riemann surface and only analytically continued into Lorentzian signature when the Riemann surface develops a nodal singularity, in other words, as we approach the boundary of the moduli space of punctured Riemann surfaces. The physical nature of these singularities depends on the type of degeneration, but the singularity of interest here arises from a non-separating degeneration. This node can be conformally mapped to an infinitely long tube, which is interpreted as a free Lorentzian string going on-shell. 

The real issue then becomes how to connect the two Lorentzian worldsheets with oppositely oriented analytic continuations. From the arguments presented above, we know that the worldsheet must degenerate in the wedge region and this degeneration suggests that we are working in the boundary of the moduli space of Riemann surfaces; these sort of spaces are naturally described by tropical geometry. Consequently, one is led to the conclusion that in order to construct the wedge region, one must investigate the behavior of analytically continued tropical worldsheets. These analytically continued tropical worldsheets were constructed in \cite{trsm} and were shown to have a non-relativistic worldsheet foliation structure as suggested by the large $N$ arguments. In particular, the tropical worldsheets appear to see the target space geometry as also being tropicalized, and hence, one is guided into asking questions surrounding the description of Hilbert space of states in these analytically continued tropical limits. We present partial answers to these questions through a tropicalization of the methods of brane quantization.

This paper is structured as follows: In section \secref{sec:BraneQuant}, we will review the basic elements of brane quantization and demonstrate how both Lagrangian and coisotropic A-branes can be used to calculate some simple invariants such as the Hilbert space dimension. In \secref{sec:TropReview}, we will review the construction of tropological sigma models with boundaries \cite{Horava:1993ts, Witten:1992fb, Albrychiewicz:2024tqe}. We will discuss how to explicitly construct Lagrangian and coisotropic A-branes in the particular case of a hyper-Kähler manifold.  In \secref{sec:TropTorus}, we will discuss an example of tropical brane quantization when replacing the topological A models with their tropical analogs and show how the Hilbert space dimensions constructed from the quantization of symplectic geometries get deformed into Hilbert space dimensions for the quantization of tropical geometries. We will do this for a simple Calabi-Yau geometry such as the cotangent bundle of $T^2$. In \secref{sec:Future}, we will discuss open questions surrounding the general construction of tropical brane quantization and several possible future research avenues.

\section{Elements of Brane Quantization}
\label{sec:BraneQuant}
We begin by reviewing the foundations of brane quantization and outlining the core elements necessary for formulating a prescription for tropical brane quantization. For additional details, we recommend the original work \cite{Gukov:2008ve} and further extensions in \cite{Gukov:2010sw, Gaiotto:2019oey}. 

In elementary treatments of quantization, one commonly encounters an inherent operator ordering ambiguity. This ambiguity often leads to the formulation of multiple, potentially inequivalent quantum systems. Over the past few decades, extensive investigations have been conducted to refine the precise definition of quantization, with the aim of understanding the conditions under which distinct quantization of the same classical system arise. This line of research has led to the development of well-refined frameworks for quantization, including geometric quantization \cite{kostant1972line, souriau1966quantification}, stochastic quantization \cite{nelson1966derivation, fenyes1952wahrscheinlichkeitstheoretische, de1967simple}, and related approaches such as deformation quantization \cite{Kontsevich:1997vb}.

Relevant to us, brane quantization provides an alternative framework for quantizing classical systems, offering a potential avenue for understanding the mathematical structures that give rise to equivalent quantizations. To be more precise, quantization typically refers to the procedure of taking a classical system described by a symplectic manifold $(M, \omega)$ and constructing an associated Hilbert space $\mathcal{H}$ along with an algebra of quantum operators $\mathcal{A}_{\mathcal{H}}$ that act on $\mathcal{H}$. Additionally, the symmetries of the symplectic geometry must be mapped to symmetries of the Hilbert space. In the absence of additional group symmetries acting on the symplectic manifold $M$, the natural symmetries of the classical system are given by the group of symplectomorphisms, while the symmetries of the Hilbert space correspond to unitary automorphisms of the algebra of quantum operators. This mapping between classical and quantum symmetries is generally not canonical and requires additional data. It is precisely at this stage that the operator ordering ambiguity arises, highlighting the fundamental question of when different quantization procedures yield inequivalent quantum mechanical systems.

There exist multiple distinct starting perspectives to brane quantization; here, we adopt a bottom-up construction. We begin by equipping the symplectic manifold $M$ with a Hermitian line bundle $\mathcal{L} \to M$ endowed with a compatible unitary connection whose curvature is $\omega$. This is known as the prequantum line bundle. In order to associate a topological A-brane to this symplectic geometry, we must extend the space by an additional symplectic form associated with the A-model. This is implemented by complexifying the symplectic manifold. By complexification, we mean an extension of $M$ to a complex symplectic manifold $\mathcal{M}$, equipped with an antiholomorphic involution $\tau$ such that $M$ is realized as a component of the fixed-point set of $\tau$. In many cases, this involution corresponds to complex conjugation of the complex coordinates that parametrize an affine or, more generally, an algebraic variety.

We require that this complexification extends all bundles originally defined on $M$. In particular, the symplectic form extends to a non-degenerate holomorphic symplectic two form $\Omega$ satisfying a compatibility condition under the involution, 
\begin{align}
    \tau^{\star} \Omega = \bar{\Omega}.
\end{align}
On the fixed-point set, this reduces to the condition that the real part of $\Omega$ restricts to the original symplectic form, 
\begin{align}
\operatorname{Re} \Omega |_{M} = \omega,
\end{align}
while the imaginary part vanishes, 
\begin{align}
    \operatorname{Im} \Omega |_{M} = 0.
\end{align}
Using the holomorphic symplectic structure, we extend the prequantum line bundle by equipping it with a connection whose curvature is now given by $\operatorname{Re} \Omega$. With this setup, we now ask under what conditions the complexified space $\CM$ supports a consistent topological A-model whose target space is  $\CM$. In essence, this requires the absence of quantum anomalies. In traditional brane quantization, a consistent topological A-model is achieved if the sigma model target space, namely $\CM$, admits a hyper-K\"ahler metric that is compatible with the holomorphic symplectic form $\Omega$ \cite{Gukov:2008ve}.  Consequently, we will probe tropical brane quantization in the simple example where the complexified target space  $T^\star T^2$ which can be endowed with hyper-K\"ahler structure .

In analogy with physical string theory, one can formulate topological A-models on Riemann surfaces with boundaries \cite{Witten:1992fb, Horava:1993ts}. In this case, one imposes boundary conditions that preserve $\CN=2$ superconformal symmetry, leading to the construction of topological analogs of D-branes, known as A-branes. Initially, these branes were considered to be supported solely on rank-1 Lagrangian submanifolds -- and hence termed Lagrangian A-branes -- which are necessarily middle-dimensional in $\CM$. In addition, Lagrangian A-branes are equipped with a unitary line bundle endowed with a flat connection. For notational convenience, we denote these Lagrangian A-branes by $\CB'$.

In \cite{Kapustin:2001ij}, it was shown that the A-model can admit A-branes whose support, denoted by $N$, has dimension exceeding one-half that of $\mathcal{M}$. In this case, the brane is supported on a rank-$1$ coisotropic submanifold $N$, which is equipped with a symplectic form $\omega_N$. Moreover, the brane carries a nontrivial Chan-Paton bundle, defined as the unitary line bundle on $N$ that encodes the gauge degrees of freedom associated with open strings ending on the brane. This bundle is endowed with a connection whose curvature is denoted by $F$. To ensure that the Chan--Paton bundle remains of rank one, one must impose the condition
\begin{align}
    \left(\omega_N^{-1} F\right)^2 = -1.
\end{align}
Following \cite{Kapustin:2006pk}, we refer to the coisotropic brane with curvature $F = \operatorname{Re}\Omega$ as the \emph{canonical} coisotropic brane, and denote it by $\mathcal{B}_{cc}$.

We now define the Hilbert space arising from the brane quantization of the symplectic manifold \(M\) by choosing a pair of A-branes i.e. a space-filling canonical coisotropic A-brane $\mathcal{B}_{cc}$ and a Lagrangian A-brane $\mathcal{B}$. In this construction, the Hilbert space
\begin{align}
\mathcal{H} = \mathrm{Hom}(\mathcal{B}_{cc},\mathcal{B}')
\end{align}
is identified with the space of open string states stretching from the canonical coisotropic brane \(\mathcal{B}_{cc}\) to the Lagrangian brane \(\mathcal{B}'\). Equivalently, in the language of the Fukaya category of the A‑model, \(\mathcal{H}\) is the space of morphisms between \(\mathcal{B}_{cc}\) and \(\mathcal{B}'\). In this procedure, the constructed Hilbert space depends only on the choice of complexification \(\mathcal{M}\), the prequantum line bundle \(\mathcal{L}\), and the data specifying \(\mathcal{B}'\); in particular, the resulting A‑model is independent of the choice of hyper-K\"ahler polarization. Moreover, the algebra of operators acting on \(\mathcal{H}\) is given by the space of \((\mathcal{B}_{cc},\mathcal{B}_{cc})\) open string states -- that is, by the algebra of operators inserted along the boundary of A‑model worldsheets ending on the canonical coisotropic brane.

It was observed in \cite{Gukov:2010sw} that the construction of the Hilbert space can be streamlined by utilizing mirror symmetry. In this approach, the space of morphisms between the relevant branes is mapped to the Ext-group of the dual branes. Consequently, the dimension of the Hilbert space can be computed using the Grothendieck-Riemann-Roch theorem.

In particular, if the base manifold under consideration is a \(2n\)-torus $T^{2n}$, the expression for the dimension simplifies to
\begin{align}
    \dim \mathcal{H} = \int_{T^{2n}} \frac{F^n}{n!},
\end{align}
where \(F\) denotes the curvature two-form associated with the corresponding line bundle. In the cases where the phase space is constrained or where additional group actions act on the symplectic manifold, one must impose additional equivariance constraints. While we do not make a general statement on the treatment of such constrained phase spaces, we highlight that a key motivation for exploring examples of tropical brane quantization is to understand how the constrained dynamics of tropical field theories \cite{trsm} influence this procedure.

\section{Tropological Sigma Models with Boundaries}
\label{sec:TropReview}

As we have discussed in the introduction, we are interested in the Hilbert spaces which can be constructed from tropological sigma models and their associated tropical geometries. Consequently, in this section, we will begin by reviewing the fundamentals of tropical geometry and how they can be implemented into quantum field theory. For additional motivations, see extensive review in \cite{trsm}.

Tropical geometry \cite{msintro, rau, mikhalkinrau, litvinov} starts by deforming the standard operations of addition and multiplication on the real number fields and then taking a singular limit. For two real numbers, $a,b \in \mathbb{R}$, the tropical addition and multiplication are 
\begin{align}
 a \odot b &=\lim_{\hbar \rightarrow 0} \hbar \log \left\{e^{(a+b) / \hbar}\right\}=a+b, \\ 
 a \oplus b &=\lim_{\hbar \rightarrow 0} \hbar \log \left\{e^{a / \hbar}+e^{b / \hbar}\right\} = \begin{cases}a, & \text { if } a>b, \\
b, & \text { if } b>a.\end{cases}
\end{align}
From the limit, one can see that multiplication in the tropical numbers corresponds to addition in the reals and addition in the tropical numbers corresponds to the maximum between the two real numbers i.e. : $a \odot b=a+b$ and $a \oplus b=\max (a, b)$. The tropical zero is played by $-\infty$. The tropical numbers $\mathbb{R}_{\hbar=0} \equiv \mathbb{T}$ now satisfy the axioms of semi-field. The corresponding limit is known as the Maslov dequantization limit \cite{msintro, litvinov}. The Maslov dequantization limit can be extended to complex manifolds by choosing complex local coordinates $Z^I$ and then taking limits of the logarithmic coordinates $ X^I \sim \log \left|Z^I\right|$. Notice that in this approach, all the phases of the complex coordinate are forgotten and what is leftover after the limit is a real algebraic geometry. This procedure yields tropical varieties that are polyhedral complexes encoding combinatorial data about the original algebraic varieties. 

Algebraic varieties are generically defined in terms of zero sets of systems of polynomials. Applying Maslov dequantization limits to a polynomial in a real variable $x$
\begin{align}
p(x)=a_1 x^{k_1}+a_2 x^{k_2}+\ldots+a_n x^{k_n}
\end{align}
where $k_1, ... k_n$ are positive integers, the Maslov dequantization limit yields a tropical polynomial of the form
\begin{align}
\operatorname{Trop}p(x)=\max \left\{a_1+k_1 x, a_2+k_2 x, \ldots, a_n+k_n x\right\}
\end{align}
In essence, a nonlinear structure in complex geometry can be transformed into a piece-wise linear structure within tropical geometry.  With this tropicalized polynomial, one  might be tempted to define tropical varieties in terms of zero sets of this tropical polynomial, however, the zero of tropical numbers is represented by $-\infty$ and consequently can lead to uninterpretable equations. Hence, in mathematics, one constructs tropical varieties in an indirect way by looking at all points where the tropical polynomial is no longer differentiable. Unfortunately, it can be quite cumbersome to implement this for physical applications since, generically, we want to investigate equations within path integrals of the form $p(x)=0$. In \cite{{trsm}}, it was discussed that the way around this is through a generalization of the field axioms where we allow the field operations to occasionally be multi-valued i.e. interpreting the tropical semifield in terms of hyperfields \cite{virohyper, viro}. In this case, one is able to interpret the Maslov dequantization/tropical limit of a complex geometry by deforming the underlying field $\mathbb{C}$.  For $z\in \mathbb{C}$ , the Maslov dequantization has the following parametrization
\begin{equation}
\label{eqn:Maslov}
\operatorname{Trop_{\hbar}}(z)=\left\{\begin{array}{lr}
|z|^{1 / \hbar} \frac{z}{|z|}, & \text { if } z \neq 0 \\
0 & \text { if } z=0
\end{array}\right.
\end{equation}
In polar coordinates $(r,\theta)$, this parametrization can be simplified as 
\begin{equation}
z=e^{r / \hbar+i \theta} .
\end{equation}

In the limit as $\hbar \rightarrow 0$, we find that we can recover the usual properties of tropical geometry. However, instead of dealing with real algebraic varieties, the tropical geometry is now modeled in terms of a multi-foliated complex geometry that retains the original manifold dimension of the initial complex variety.  This foliation admits possible singularities. This is a distinct perspective on tropical geometry as opposed to the mathematical literature because we choose to actively retain the phases of the complex numbers. Taking a tropical limit generically does not only affect the original manifold but also affects all bundles constructed upon the original manifold. 

In particular, taking the tropical limit of a Riemann surface $\Sigma$ results in the complex structure being deformed to a Jordan structure, which is a nilpotent endomorphism of the tangent bundle i.e., for $\varepsilon: T\Sigma \rightarrow T\Sigma$, we have
\begin{equation}
\label{eqn:JordanStr}
\varepsilon^2=0.
\end{equation}
It is precisely this Jordan structure that gives rise to a fiberwise filtration of vector spaces, which defines a one dimensional integrable distribution of the tangent bundle, that allows one to represent a real one-dimensional tropical geometry as a complex one-dimensional foliated Riemann surface. In the case of higher dimensional complex geometries, the Jordan structures can give rise to far more complicated foliations and give rise to generalized tropical geometries, including odd-dimensional ones associated with contact geometry.

One might be tempted to take these sort of limits on other geometric objects that admit local coordinate descriptions.  If one has a local trivialization of a vector bundle over some complex variety, one may be tempted to take Maslov dequantization limits of not only the base coordinates and likewise, not only the fiber coordinates, but instead a combination of the base and fiber coordinates. It is not obvious what sort of mathematical object this yields.  When this limit results in an tropical variety, we say that we have \textit{tropicalized} the geometry.  

\subsection{Tropological Sigma Models}
\label{sec:ReviewTropSigma}

This interpretation of tropicalization allows us to construct conventional path integrals out of complex numbers instead of being forced to construct a path integral made out of multi-valued hyperfields. In this case, the multivaluedness appears as additional gauge invariances within the path integral. These gauge invariances are anisotropic in nature and, hence, pre-tropicalization, we label field theories as relativistic, and post-tropicalization, as non-relativistic and/or tropical\footnote{We use this terminology because we begin the construction of tropological sigma models through relativistic sigma models. In principle, there is nothing stopping you from a taking a tropical limit of an originally non-relativistic field theory.}. Using these insights, one is then able to construct tropical sigma models by either taking direct Maslov dequantization limits of the worldsheet actions or in the case of topological sigma models, by taking tropical limits of the appropriate localization equations and then constructing a cohomological field theory through standard BRST cohomology methods.  For the purposes of brane quantization, we will be interested in finding the tropical analogs of Lagrangian A-branes and coistropic A-branes and thus we will quickly review the construction of tropical topological sigma A models i.e., \textit{tropological sigma models} to establish notation.

We begin with a standard, relativistic 2-dimensional supersymmetric sigma model, which is a field theory, that describes the dynamics of maps $\Phi$ from a source space  $\Sigma$ to a target space $M$.  The source space in this context is a 1 complex dimensional manifold known as the worldsheet that is equipped with a complex structure $\hat{\varepsilon}$, where we use a hat to denote geometric structures before taking the tropical limit.  The target space $M$ is a complex manifold of arbitrary complex dimension equipped with an almost complex structure $\hat{J}$ .  The fields $\Phi$ come along with superpartners $\psi$ which are worldsheet 1-forms valued in the pullback of the odd tangent bundle of $M$.  The topological A-model is then constructed by a topological twist such that the path integral of the A-model localizes onto psuedoholomorphic maps 
\begin{equation}
\widehat{J} \circ d \Phi=d \Phi \circ \widehat{\varepsilon}.
\end{equation}
We can establish local coordinates $\sigma^{\alpha}$, $\alpha \in\{1,2\}$ on the source space $\Sigma$ and local coordinates $Y^i$, $i \in \{1,2,...,2n\}$ on the target space $M$ and locally write down these localization equations as
\begin{equation}
\label{eqn:LocEqn}
\widehat{E}_\alpha{ }^i=\widehat{\varepsilon}_\alpha^\beta \partial_\beta Y^i-\widehat{J}_j^i \partial_\alpha Y^j=0
\end{equation}
We are able to take the tropical limit of $\widehat{E}_\alpha{ }^i$ by noting that the field variables in this case can also be interpreted as embedding coordinates on the target space, we write 
\begin{equation}
z=\exp \left\{\frac{r}{\hbar}+i \theta\right\}, \quad Z=\exp \left\{\frac{X}{\hbar}+i \Theta\right\}
\end{equation}
Taking the tropical limit of \eqref{eqn:LocEqn} formally replaces the complex structures $\hat{\varepsilon}$, and $\hat{J}$ with their tropical analogs $\varepsilon, J$,  which are known as Jordan structures. These Jordan structures are characterized by their nilpotency, as in \eqref{eqn:JordanStr}, i.e.,
\begin{equation}
J^2=0, \quad \varepsilon^2=0.
\end{equation}

One finds that the symmetries that preserve the Jordan structures can be interpreted as a nonrelativistic conformal symmetry, and the worldsheet diffeomorphisms reduce down to worldsheet foliation-preserving diffeomorphism given by
\begin{equation}
\begin{aligned}
& \widetilde{r}=\widetilde{r}(r), \\
& \widetilde{\theta}=\widetilde{\theta}_0(r)+\theta \partial_r \widetilde{r}(r) .
\end{aligned}
\end{equation}
These foliation-preserving diffeomorphisms also preserve the form of the worldsheet Jordan structure. The tropical localization equations are then
\begin{equation}
E_r^X=\partial_\theta X=0, \quad E_r^{\Theta}=\partial_\theta \Theta-\partial_r X=0, \quad E_\theta^X=0, \quad E_\theta^{\Theta}=-\partial_\theta X=0
\end{equation}
One finds that the localization equations enjoy an additional gauge symmetry known as an $\alpha$ symmetry.  Infinitesimally, the $\alpha$ symmetry is written down as
\begin{equation}
\begin{aligned}
\label{eqn:AlphaSym}
\delta X &= \alpha_1(r), \\
\delta \Theta &= \alpha_0(r)+\theta \partial_r \alpha_1(r).
\end{aligned}
\end{equation}

In order to construct a Lagrangian representative in terms of BRST cohomology, we postulate that the tropical fields X and $\Theta$ transform as scalars under foliation-preserving diffeomorphisms, thus
\begin{equation}
\begin{aligned}
\label{eqn:TropDiff}
\delta X & =f(r) \partial_r X+\left(F(r)+\theta \partial_r f(r)\right) \partial_\theta X \\
\delta \Theta & =f(r) \partial_r \Theta+\left(F(r)+\theta \partial_r f(r)\right) \partial_\theta \Theta.
\end{aligned}
\end{equation}
One might be tempted to use the worldsheet metric, in order to construct a Lagrangian action. But upon tropicalization, one finds that we have a degenerate worldsheet metric and the corresponding Riemannian volume form vanishes. Consequently, we will work with tensor densities to regulate the divergences that arise in the tropical limit. We review the main steps, referring to the full discussion in \cite{trsm}.

We introduce the following tensor densities:  $\CB^{\alpha}_{\;\;i}$, a bosonic auxiliary field, together with an antighost superpartner $\chi^{\alpha}_{\;\;i}$, which are put together into an antighost BRST multiplet
\begin{align}
    \{Q,\chi^{\alpha}_{\;\;i}\}&=\CB^{\alpha}_{\;\;i}, \\
    [Q,\CB^{\alpha}_{\;\;i}]&=0,
\end{align}
with a nilpotent graded differential, known as the  BRST charge $Q$. This BRST charge, when acted on fields $X^i$, reflects an underlying topological symmetry $[Q,X^i]=\psi^i$, where $\psi^i$ are Grassmann odd ghost fields. 

The BRST invariant action can then be written as
\begin{align}
    \int_\Sigma d^2\sigma \CB^{\alpha}_{\;\;i}E_{\alpha}^{\;\;i},
\end{align}
up to an addition of BRST exact terms that preserve both the non-degeneracy and the BRST cohomology class of the path integral integrand. The auxiliary field $\CB^{\alpha}_{\;\;i}$ is, by construction, gauge invariant. We are able to use the $\alpha$ symmetry \eqref{eqn:AlphaSym} to set half of the components of $\CB^{\alpha}_{\;\;i}$ to zero.  In particular, we set
\begin{align}
    \CB^{\alpha}_{\;\;X}=0.
\end{align}
We denote the leftover components as
\begin{align}
    \CB^{r}_{\;\;\Theta}=B, \quad \CB^{\theta}_{\;\;\Theta}=-\beta.
\end{align}
In this gauge, and after integrating out a term quadratic in $B$, the bosonic part of the action is
\begin{align}
\label{eqn:TropAction}
    S=\int_{\Sigma} dr d\theta\left\{\frac{1}{2}(\partial_\theta\Theta-\partial_r X)^2+\beta\partial_\theta X\right\}.
\end{align}
One can notice that we did not integrate out the auxiliary field $\beta$. The reason is that we cannot add a $\beta^2$ term that is consistent with the anisotropic conformal invariance \eqref{eqn:TropDiff}.

\subsection{Tropical Hyper-K\"ahler Manifolds }
\label{sec:TropHyperKahler}

It is well known that Lagrangian A-branes and coisotropic A-branes can arise as supersymmetry preserving boundary conditions on worldsheets; we will take this worldsheet perspective and begin by considering the tropological sigma A-model on a foliated Riemann surface $\Sigma$ with boundary $\partial \Sigma$.  Since we are only interested in seeing the basic properties of tropical brane quantization, we focus on the particular case where the the target space of the sigma model is tropicalization of the complexified symplectic manifold $\CM = T^\star T^2$.  For the case of $T^\star T^2$,  it is clear what sort of tropical geometry one obtains after the Maslov dequantization limit \eqref{eqn:Maslov} of the worldsheet. However, there are generically many inequivalent tropicalizations that one can take of the target space since it is not a 1 complex dimensional manifold.

To construct an interesting tropicalization of the target space, we aim to preserve essential structures present in the standard framework of brane quantization. In particular, a topological A-model remains anomaly-free provided that the first Chern class of the target space, $c_1(\mathcal{M})$, vanishes. If, in addition to this topological condition, the space is Ricci-flat, then it qualifies as a Calabi–Yau manifold. Notably, hyper-K\"ahler manifolds, such as $T^\star T^2$, serve as convenient examples of Calabi–Yau spaces. Motivated by these considerations, we seek to define a tropical limit that retains aspects of the hyper-K\"ahler condition.

Before we attempt to construct a tropical hyper-K\"ahler condition, we will start by investigating how one can extract interesting tropical limits from a 1 complex dimensional K\"ahler manifold $Y$, where the geometric objects of interest are a metric tensor $g$, a complex structure $\varepsilon$ and a symplectic form $\omega$. For $u,v \in \Gamma(TY)$,  the K\"ahler compatibility condition $g(u, v)=\omega(\varepsilon(u),  v)$ can be stated without reference to the sections as
\begin{align}
\hat{g}=(\hat{\varepsilon}\otimes \mathbb{I})^{*}\hat{\omega},
\end{align}
where we have defined $\varepsilon\otimes \mathbb{I}$ as an operator that acts on $TY \times TY$ by a pullback through $\omega$.  $\mathbb{I}$ is the identity operator on $TY$ and the hats, as defined before, refer to geometric objects prior to the tropical limit. Recall that we can show through the Maslov dequantization limit that the complex structure reduces down to a Jordan structure that satisfies $\varepsilon^2=0$. In the adapted coordinate system $(r,\theta)$, one also obtains the following form for the tropical metric and Jordan structure
\begin{align}
    g=dr \otimes dr, \; \varepsilon= \frac{\partial}{\partial r} \otimes d \theta.
\end{align}
In particular, one finds that the symplectic form does not change in the adapted coordinates  $\omega=dr\wedge d\theta$. In passing, we also mention that one can show that that the tropical metric tensor $g$ now satisfies a ``mutual invisibility" condition with the tropical inverse metric tensor $h$ instead of the usual inverse law i.e.,
\begin{equation}
g_{\alpha \beta} h^{\beta \gamma}=0, \quad h^{\alpha \beta} g_{\beta \gamma}=0 .
\end{equation}

Moving onto the case where we now have a hyper-K\"ahler manifold, we now have an algebra of integrable almost complex structures $I,J,K$ that individually satisfy the K\"ahler compatibility condition and additionally satisfy the quaternionic relations $I^2=J^2=K^2=IJK=-1$.  One can realize an $\su$ matrix representation of the complex structures with commutation relations
\begin{align}
    [I,J]=2K, \quad [J,K]=2I, \quad [K,I]=2J.
\end{align}
We can investigate potential contractions of the $\su$ algebra by employing the Wigner-\.In\"on\"u \cite{inonu1952} contraction. This algebra can be contracted in three distinct ways, resulting in $\mathfrak{a}_3, \mathfrak{h}_3$ and $\mathfrak{iso}(2)$ \cite{goze2019coadjoint}. Here, $\mathfrak{a}_3$ is the maximally abelian algebra of $\MR^3$, $\mathfrak{h}_3$ is the Heisenberg algebra, which is a non-abelian nilpotent Lie algebra that appears frequently in canonical quantization, and $\mathfrak{iso}(2)$ is the Lie algebra of the Euclidean group $\mathrm{E}(2) \cong S O(2) \ltimes \mathbb{R}^2$ which describes isometries of Euclidean plane.

We employ the  Wigner-\.In\"on\"u contraction by  rescaling the complex structures $I,J,K$ by the same $\epsilon$ factor: $(I,J,K)\rightarrow (\epsilon i, \epsilon j, \epsilon k)$, where the lowercase $i , j, k$ denote the rescaled structures. Taking the limit $\epsilon \rightarrow 0$, we see that the Lie algebra trivializes with all commutators of the generators vanishing. We identify this as $\mathfrak{a}_3$. If, we instead rescale by sending $(I,J,K)\rightarrow (\sqrt{\epsilon} i,\sqrt{\epsilon} j, \epsilon k)$, we obtain the Heisenberg algebra $\mathfrak{h}_3$ 
\begin{align}
    [i,j]=2k, \quad [j,k]=0, \quad [k,i]=0.
\end{align}

Finally,  performing the rescaling $(I,J,K)\rightarrow (\sqrt{\epsilon} i,\sqrt{\epsilon} j, k)$, gives the $\mathfrak{iso}(2)$ algebra
\begin{align}
\label{eqn:SU2toIso2}
    [i,j]=0, \quad [j,k]=2i, \quad [k,i]=2j.
\end{align}

For the purposes of brane quantization, we will mostly focus on the last contraction that yields $\mathfrak{iso}(2)$. This contraction naturally appears when we tropicalize the hyper-K\"ahler manifold.  Using local coordinates, one is able to show that in the tropical limit, two of the complex structures reduce down to Jordan structures and one remains a conventional complex structure
\begin{align}
    i^2=0, \quad j^2=0, \quad k^2=-1.
\end{align}
The Wigner-\.In\"on\"u contraction along with the tropical limit have implications on the results of the brane quantization. For ease of notation, we once again relabel $(i,j,k)\rightarrow (I,J,K)$. We call the resulting geometry, a \textit{tropical hyper-K\"ahler manifold.}

As explained in \secref{sec:BraneQuant}, on a complexified manifold $\CM$, we may select a holomorphic two form $\Omega$ that is of type $(2,0)$ with respect to $J$. Unlike \cite{Gukov:2008ve}, where more general scenarios are considered,  we restrict our attention to the case where  $\CM$ has a hyper-K\"ahler structure such that $J$ satisfies quaternionic relations. $\Omega$ can be decomposed into its real and imaginary parts
\begin{align}
\label{eqn:OmegaDecomp}
    \Omega = \omega_I+i\omega_K,
\end{align}
where one can show by pullback that $(J \otimes \mathbb{I})^* \Omega = i\Omega$ , in particular
\begin{align}
\label{eqn:SymCM}
    (J \otimes \mathbb{I})^*\omega_I=-\omega_K, \quad (J \otimes \mathbb{I})^*\omega_K=\omega_I.
\end{align}

As discussed in \secref{sec:BraneQuant}, it is possible to have a coisotropic A-brane supported on the entire manifold $\CM$.  The aforementioned coistropic A-brane is equipped with a rank 1 Chan-Paton bundle which is a unitary line bundle $\CL$  with a non-vanishing curvature  $F$.  In order to preserve the supersymmetry that defines the A-brane, one finds that the necessary condition is that we can construct a complex structure from the curvature as $J=\omega_K^{-1}F$.  From \eqref{eqn:SymCM}, we can satisfy this by setting $F=\omega_I$ and inverting the equation. Since $\omega_K$ is of type $(1,1)$ with respect to $K$, we can construct an A model on $\CM$ with the metric given by $g=(K \otimes \mathbb{I})^* \omega_K$. These statements get appropriately modified under the tropicalization of $I,J,K$.

One is able to choose the tropicalization in such a way that two of the original complex structures become Jordan structures and the last one remains a conventional complex structure. Keeping the roles of $I,J,K$ fixed in this process, so that $J$ is the complex structure with respect to which we construct $\Omega$, we will consider two different cases. 

In the first case, we keep $K$ as a complex structure and both $I$ and $J$ become Jordan structures. In this case, the relativistic localization equations are preserved, however the metric degenerates because now $\omega_K$ is degenerate. The symplectic two form $\omega_I$ of the manifold $M$ we started with is unmodified, however, we will find that its integral over $M$ will give different a Hilbert space dimension than in the relativistic case.  

In the second case, we keep $I$ as a complex structure and both $J,K$ become Jordan structures. Now, the associated A model is a tropological A model with a Jordan structure $K$ and degenerate metric. Since the symplectic two form $\omega_I$ is degenerate, we find that the dimension of the Hilbert space vanishes in this case.

\subsection{Tropical Lagrangian Branes and Coistropic Branes}
Now we present the explicit construction of the tropical Lagrangian A-branes and tropical coisotropic A-branes for the case where the target is a tropical hyper-K\"ahler manifold. As mentioned in the previous section, we begin with a foliated Riemann surface $\Sigma$ with boundary $\partial\Sigma $ whose local coordinate patches are characterized by foliation-preserving diffeomorphisms and whose target is $\CM=T^*T^2$. The localization equation of the tropical A model are written using the worldsheet Jordan structure $\varepsilon$ and the target space Jordan structure $K$ in local coordinates as
\begin{align}
\label{eqn:TropEqnHyperKahler}
    E_{\alpha}^{\; i}\equiv \varepsilon_{\alpha}^\beta\partial_\beta Y^i-K_{i}^{\; j}\partial_\alpha Y^j=0,
\end{align}
Using adapted local coordinates on the tropicalized worldsheet, we parametrize the target space base manifold $T^2$ by $(\Theta, \Psi)$ and fibers by $(X,Y)$. In these coordinates 
\begin{align}
    K=\frac{\partial}{\partial X}\otimes d\Psi-\frac{\partial}{\partial Y}\otimes d\Theta,
\end{align}
which we construct with more details in \secref{sec:TropTorus}. 
Now, the localization equations \eqref{eqn:TropEqnHyperKahler} are explicitly
\begin{align}
\label{eqn:TropLocEqn}
    E_{r}^{\;\; X}&=\partial_\theta X, \quad E_{r}^{\;\; \Theta}=\partial_\theta\Theta+\partial_r Y, \quad E_{r}^{\;\; Y}=\partial_\theta Y, \quad E_{r}^{\;\; \Psi}=\partial_\theta\Psi-\partial_r X \\
    E_{\theta}^{\;\; X}&=0, \quad E_{\theta}^{\;\; \Theta}=\partial_\theta X, \quad E_{\theta}^{\;\; Y}=0, \quad E_{\theta}^{\;\; \Psi}=-\partial_\theta X.
\end{align}
The solution to these equations are
\begin{align}
    X(r,\theta)&=X_0(r), \\
    Y(r, \theta) &= Y_0(r),\\
    \Theta(r,\theta)&=\Theta_0(r) - \theta\partial_r Y_0(r), \\
    \Psi(r,\theta)&=\Psi_0(r) + \theta \partial_r X_0(r).
\end{align}
With the localization equations we can use cohomological BRST quantization to construct a Lagrangian path integral in the manner described in \cite{trsm}.   

Introducing the auxiliary field $\CB^\alpha_{\;\; i}$, the action may be written as
\begin{align}
    S=\int_{\Sigma} d^2\sigma \left(\CB^\alpha_{\;\; i}E_\alpha^{\;\; i}-\frac{1}{2}\gamma_{\alpha\beta}h^{ij}\CB^{\alpha}_{\;\; i}\CB^{\beta}_{\;\; j}\right),
\end{align}
where $\gamma_{\alpha\beta}$ is the tropical worldsheet metric and $h^{ij}$ is the tropical inverse metric tensor on $\hat{M}$. There is a redundancy in $\CB^\alpha_{\;\; i}$, which we fix by setting:
\begin{align}
    \CB^\alpha_{\;\; X}=0, \quad \CB^\alpha_{\;\; Y}=0,
\end{align}
as discussed in \cite{trsm},  this is done at the cost of abandoning the full covariance in the target space. We will denote leftover components as:
\begin{align}
    \CB^r_{\;\; \Theta}=B_1, \quad \CB^r_{\;\; \Psi}=B_2, \quad \CB^\theta_{\;\; \Theta}=\beta_1, \quad \CB^\theta_{\;\; \Psi}=-\beta_2.
\end{align}
Since $h^{\theta\theta}$ and $h^{\Psi\Psi}$ are non-zero, we can integrate out $B_1$ and $B_2$. The final action is then
\begin{align}
\label{eqn:TropTorusAction}
    \int_{\Sigma} dr d\theta\left\{\frac{1}{2}\left(\partial_\theta\Theta+\partial_r Y\right)^2+\frac{1}{2}\left(\partial_\theta\Psi-\partial_r X\right)^2+\beta_1\partial_\theta Y+\beta_2\partial_\theta X\right\}.
\end{align}
and the classical equations of motion are locally solved by
\begin{align*}
    Y(r,\theta)&=Y_0(r), \\
    X(r,\theta)&=X_0(r), \\
    \Theta(r,\theta)&=\Theta_0(r)-\theta(\partial_r Y_0(r)-C(r)), \\
    \Psi(r,\theta)&=\Psi_0(r)+\theta(\partial_r X_0(r)+D(r)), \\
    \beta_1(r,\theta)&=-\theta\partial_r C(r)+\beta_{1,0}(r), \\
    \beta_2(r,\theta)&=-\theta\partial_r D(r)+\beta_{2,0}(r), 
\end{align*}
where the functions $C(r)$ and $D(r)$ are arbitrary projectable and differentiable functions on the foliation. 

Now, we will consider the boundary conditions on $\partial\Sigma$, which we represent as the fixed point set of a worldsheet involution given by complex conjugation i.e. the points where $z=\bar{z}$. In the adapted coordinate system
\begin{align}
    z=\exp\left\{\frac{r}{\hbar}+i\theta\right\}, \; \bar{z}=\exp\left\{\frac{r}{\hbar}-i\theta\right\},
\end{align}
the worldsheet boundary is then given by the points that correspond to $\theta=\{0, \pi\}$. We will begin with the Lagrangian A brane where we have a vector bundle on the worldsheet with zero curvature. Taking the constructed Lagrangian action \eqref{eqn:TropTorusAction}, we vary the fields and obtain the following boundary terms
\begin{align}
    \int_{\partial\Sigma} dr\left\{\left(\partial_\theta\Theta+\partial_r Y\right)\delta\Theta |_{\theta=0,\pi}+\left(\partial_\theta\Psi-\partial_r X\right)\delta\Psi |_{\theta=0,\pi}+\beta_1\delta Y|_{\theta=0,\pi}+\beta_2\delta X|_{\theta=0,\pi}\right\}.
\end{align}
This gives two types of boundary conditions. The first set can be identified with Neumann boundary conditions
\begin{align}
    \partial_\theta\Theta+\partial_r Y=0, \quad \partial_\theta \Psi-\partial_r X=0, \quad \beta_1=0, \quad \beta_2=0.
\end{align}
The second one is Dirichlet boundary conditions
\begin{align}
    \partial_r X=0, \quad \partial_r\Theta=0, \quad \partial_r Y=0, \quad \partial_r \Psi=0.
\end{align}
For the Lagrangian brane the boundary conditions are mixed. One will have Neumann boundary conditions in tangent directions and Dirichlet boundary conditions in normal directions.

In order to discuss the boundary conditions that lead to a coisotropic brane, and in what follows we only consider a space filling one, we need to couple a gauge field $A$ with nonzero curvature $F$  to the boundary of the tropological sigma model. We do this by pulling back the gauge field $A$ to the boundary
\begin{align}
    \int_{\partial\Sigma}\Phi^\star(A),
\end{align}
with the caveat that it has been properly defined so that the total action is still invariant under foliation preserving diffeomorphisms.

In our coordinates system, we will pick $F=\omega_I=-dX\wedge dY$, and this choice is motivated by the discussion in \secref{sec:TropTorus}. With this $F$, the gauge field coupling gives an addition boundary term
\begin{align}
    \int_{\partial\Sigma} dr(-\partial_r X\delta Y|_{\theta=\pi}+\partial_r Y\delta X|_{\theta=\pi}).
\end{align}
For a tropical space filling coisotropic brane, the tropical boundary conditions are then given by modified Neumann boundary conditions
\begin{align}
    \partial_\theta \Theta+\partial_r Y&=0, \quad \partial_\theta \Psi-\partial_r X=0, \\ 
    \beta_1-\partial_r X&=0, \quad \beta_2+\partial_r Y=0.
\end{align}

For the purposes of brane quantization, we want to investigate solutions where we have a tropical Lagrangian A-brane at $\theta=0$ and tropical coisotropic A-brane at $\theta=\pi$.  Implementing this into the equations of motion, we obtain
\begin{align}
    \label{eqn:CoisotropicSol}
    X(r,\theta)&=x_0, \\ \nonumber
    \Theta(r, \theta)&=\Theta_0(r), \\ \nonumber
    Y(r,\theta)&=y_0, \\ \nonumber
    \Psi(r, \theta) &= \Psi_0(r), \\ \nonumber
    \beta_1(r,\theta)&=\beta_2(r,\theta)=0,
\end{align}
with $x_0, y_0$ being constant.  This is distinct from  the construction in \cite{Albrychiewicz:2024tqe}, since the tropological sigma models that were used there were the analytic continuations constructed for the sake of real-time physics; here we have focused on the construction of boundaries -- and hence branes -- for the topological case. We will proceed to providing two examples of dequantization limits of $\CM$.

\section{Examples of Tropical Brane Quantization}
\label{sec:TropTorus}

In this section, we will place hats on all objects before tropicalization and remove the hats only after we have taken the tropical limit.
Let us briefly summarize again the data we need for brane quantization \cite{Gukov:2010sw}:
\begin{itemize}
    \item We introduce a complexification of $\hat{M}=T^2$ given by cotangent bundle of the torus, which we call $\hat{\CM}=T^\star T^2=T^2\times \mathbb{R}^2$ equipped with a complex structure $\hat{J}$ and an antiholomorphic involution $\tau$, which is an automorphism on $\hat{\CM}$ such that $\hat{M}$ is a component of the fixed point set of $\tau$. We locally trivialize the the cotangent bundle by making the identification $T^{*}T^2\cong C^\times \times C^\times$.
    \item The symplectic form $\hat{\omega}$ of $\hat{M}$ is given by the restriction of the real part of the nondegenerate holomorphic two form $\hat{\Omega}$ of $\hat{\CM}$. $\hat{\Omega}$ is of type $(2,0)$ with respect to $\hat{J}$ and it satisfies $\tau^\star\hat{\Omega}=\overline{\hat\Omega}$.
    \item We extend the original prequantum line bundle $\mathcal{L}\rightarrow \hat{M}$ to the unitary line bundle $\mathcal{L}\rightarrow \hat{\CM}$ with a connection of curvature $\text{Re}\, \hat{\Omega}$.
\end{itemize}
If in this procedure, the complexified manifold $\hat{\CM}$ admits a complete hyper-K\"ahler metric and therefore an A-model, with a real symplectic form given by $\text{Im}\, \hat{\Omega}$, then one is able to construct a Hilbert space associated to $\hat{M}$ through the quantization of the topological A model.

In the case of $T^2$, the complexified manifold has a natural holomorphic symplectic two form:
\begin{align}
\label{eqn:OldHolo2Form}
    \hat{\Omega} = \frac{1}{2\pi}\frac{dZ_1}{Z_1}\wedge \frac{dZ_2}{Z_2},
\end{align}
which can be shown to be type $(2,0)$ with respect to $\hat{J}$, as required above. In this case $\hat{\CM}=T^*T^2$ is hyper-K\"ahler. We can parametrize the cotangent bundle using local coordinates $(\Theta,\Psi)$ on the base and fiber coordinates $(X,Y)$. This hyper-K\"ahler structure can then be seen explicitly by parametrizing coordinates on both copies of $C^\times$ as:
\begin{align}
\label{eqn:TargetCoords}
    Z_1=e^{X+i\Theta}, \quad Z_2=e^{Y+i\Psi},
\end{align}
then the explicit form of $\hat{\Omega}$ is: 
\begin{align}
\label{eqn:HoloTwoForm}
    \hat{\Omega}=\frac{1}{2\pi}(dX\wedge dY-d\Theta\wedge d\Psi)+\frac{i}{2\pi}(d\Theta\wedge dY+dX\wedge d\Psi)=-\hat{\omega}_I+i\hat{\omega}_K,
\end{align}
where we identified symplectic forms $\hat{\omega}_I, \hat{\omega}_K$ as real and imaginary parts of holomorphic two forms and add minus sign for $\hat{\omega}_I$ (cf. \eqref{eqn:OmegaDecomp}) to preserve chosen orientation. The third symplectic form $\hat{\omega}_J$ is then uniquely defined. We can also write an explicit form of the involution $\tau$ that preserves $T^2$ as a component of the fixed point set:
\begin{align}
    \tau(X,\Theta, Y, \Psi)\rightarrow (-X, \Theta, -Y, \Psi),
\end{align}
under which $\tau^\star(\hat{\Omega})=\overline{\hat{\Omega}}$ and $\tau^\star(\hat{\omega}_K)=-\hat{\omega}_K$, $\tau^\star(\hat{\omega}_I)=\hat{\omega}_I$. 

In local coordinates, the matrix representations of the complex structures that we use are,
\begin{align}
\label{eqn:ComplexStr}
    \hat{I}= \begin{pmatrix}
    0 & 0 & -1 & 0 \\
    0 & 0 & 0 & 1 \\
    1 & 0 & 0 & 0 \\
    0 & -1 & 0 & 0
    \end{pmatrix}, \quad
     \hat{J}= \begin{pmatrix}
    0 & 1 & 0 & 0 \\
    -1 & 0 & 0 & 0 \\
    0 & 0 & 0 & 1 \\
    0 & 0 & -1 & 0
    \end{pmatrix},
    \quad
     \hat{K}= \begin{pmatrix}
    0 & 0 & 0 & 1 \\
    0 & 0 & 1 & 0 \\
    0 & -1 & 0 & 0 \\
    -1 & 0 & 0 & 0
    \end{pmatrix}.
\end{align}
We observe that symplectic form $\hat{\Omega}$ restricted to $\hat{M}$ is simply $-d\Theta\wedge d\Psi$, which is the expected symplectic form for $T^2$ . The symplectic form for the A-model $\hat{\omega}_K$ vanishes within the fixed point set of $\tau$. In fact, since $F=\text{Re}\, \hat{\Omega}$ , we can compute the dimension of the Hilbert space via 
\begin{align}
    \dim \mathcal{H}=\frac{1}{2\pi\hbar}\int_{T^2} F = \frac{1}{\hbar}.
\end{align}
To obtain the associated A-model, we use the canonical metric tensor given by the Kähler compatibility condition $g=(K\otimes \mathbb{I})^{*}\omega_K$. 

\subsection{Tropical Limit of the Complexified Torus}
We now apply the Maslov dequantization limit \eqref{eqn:Maslov} to the cotangent bundle in such a way that we allow for mixing between the base coordinates $(\Theta,\Psi)$ and fiber coordinates $(X,Y)$. In this particular case, since we are not taking the usual Maslov dequantization limits where the base and fiber coordinates are not mixed, it is not obvious that we obtain a tropical geometry in the limit $\hbar\rightarrow 0$ . The associated deformed coordinates are then
\begin{align}
\label{eqn:TropTargetCoords}
    Z_1=e^{\frac{X}{\hbar}+i\Theta}, \quad Z_2=e^{\frac{Y}{\hbar}+i\Psi}.
\end{align}
The deformation of the coordinates modifies the holomorphic two form $\hat{\Omega}$
\begin{align}
    \hat{\Omega}=\frac{1}{2\pi}\left(\frac{1}{\hbar^2}dX\wedge dY-d\Theta\wedge d\Psi\right)+\frac{i}{2\pi}\left(\frac{1}{\hbar}d\Theta\wedge dY+\frac{1}{\hbar}dX\wedge d\Psi\right),
\end{align}
hence the real symplectic two forms $\hat{\omega}_I, \hat{\omega}_J$ and $\hat{\omega}_K$ are now deformed
\begin{align}
    \hat{\omega}_I&=\frac{1}{2\pi}\left(d\Theta\wedge d\Psi-\frac{1}{\hbar^2}dX\wedge dY\right), \\
    \hat{\omega}_J&=\frac{1}{2\pi}\left(\frac{1}{\hbar}dX\wedge d\Theta+\frac{1}{\hbar}dY\wedge d\Psi\right), \\
    \hat{\omega}_K&=\frac{1}{2\pi}\left(\frac{1}{\hbar}d\Theta\wedge dY+\frac{1}{\hbar}dX\wedge d\Psi\right)
\end{align}
The metric tensor also gets deformed as follows
\begin{align}
    \hat{g}&=\hat{g}_{\mu\nu}dx^{\mu}\otimes dx^{\nu}=\frac{1}{|Z_1|^2}dZ_1\otimes d\overline{Z}_1+\frac{1}{|Z_2|^2}dZ_2\otimes d\overline{Z}_2 \\
    &=\frac{1}{\hbar^2}dX\otimes dX+d\Theta\otimes d\Theta+\frac{1}{\hbar^2}dY\otimes dY+d\Psi\otimes d\Psi.
\end{align}
Recall that we want to preserve the hyper-K\"ahler structure on $\hat{\CM}$ in the same way as in \secref{sec:TropHyperKahler}, therefore we want to assure that relation between the complex structures, symplectic structure and metric tensor is preserved under the tropical limit:
\begin{align}
\label{eqn:HyperKahlerRelations}
    \omega_I=(I\otimes \mathbb{I})^{*}g, \quad \omega_J=(J\otimes \mathbb{I})^{*}g, \quad \omega_K=(K\otimes \mathbb{I})^{*}g,
\end{align}
The deformation of the complex structures can be written in matrix representation as
\begin{align}
    \hat{I}= \begin{pmatrix}
    0 & 0 & -1 & 0 \\
    0 & 0 & 0 & 1 \\
    1 & 0 & 0 & 0 \\
    0 & -1 & 0 & 0
    \end{pmatrix}, \quad
    \hat{J}= \frac{1}{\hbar}\begin{pmatrix}
    0 & 1 & 0 & 0 \\
    -\hbar^2 & 0 & 0 & 0 \\
    0 & 0 & 0 & 1 \\
    0 & 0 & -\hbar^2 & 0
    \end{pmatrix},
    \quad
    \hat{K}= \frac{1}{\hbar}\begin{pmatrix}
    0 & 0 & 0 & 1 \\
    0 & 0 & \hbar^2 & 0 \\
    0 & -1 & 0 & 0 \\
    -\hbar^2 & 0 & 0 & 0
    \end{pmatrix},
\end{align}
in the limit $\hbar\rightarrow 0$, they contract to:
\begin{align}
\label{eqn:TropComplexStr}
    I= \begin{pmatrix}
    0 & 0 & -1 & 0 \\
    0 & 0 & 0 & 1 \\
    1 & 0 & 0 & 0 \\
    0 & -1 & 0 & 0
    \end{pmatrix}, \quad
     J= \begin{pmatrix}
    0 & 1 & 0 & 0 \\
    0 & 0 & 0 & 0 \\
    0 & 0 & 0 & 1 \\
    0 & 0 & 0 & 0
    \end{pmatrix},
    \quad
     K= \begin{pmatrix}
    0 & 0 & 0 & 1 \\
    0 & 0 & 0 & 0 \\
    0 & -1 & 0 & 0 \\
    0 & 0 & 0 & 0
    \end{pmatrix}.
\end{align}
Comparing to \eqref{eqn:ComplexStr}, we observe that tropical limit contracts two of the complex structures into Jordan structures and one is left as a conventional complex structure. In fact, we can identify that under tropicalization the original algebra $\mathfrak{su}(2)$ of \eqref{eqn:ComplexStr} is contracted to an $\mathfrak{iso}(2)$ algebra (cf. \eqref{eqn:SU2toIso2}):
\begin{align}
    I^2=-1, \; K^2=J^2=0, \; [I,J]=2K, \; [K,I]=2J, \; [J,K]=0.
\end{align}
In the limit, one also obtains for the symplectic forms after a multiplicative renormalization by $\hbar$ :
\begin{align}
\label{eqn:TropSymp}
    \omega_I&=-\frac{1}{2\pi}\left(dX\wedge dY\right), \\
    \omega_J&=\frac{1}{2\pi}\left(dX\wedge d\Theta+dY\wedge d\Psi\right), \\
    \omega_K&=\frac{1}{2\pi}\left(d\Theta\wedge dY+dX\wedge d\Psi\right), 
\end{align}
and the tropical metric: 
\begin{align}
\label{eqn:AmodelMetric}
    g=dX\otimes dX+dY\otimes dY.
\end{align}
In this limit, we lose the Bohr-Sommerfeld quantization condition since the integral of $\omega_I$  over $M$ now vanishes.

\subsection{Different Choice of Tropical Limit}
\label{sec:DifferentQuant}
In this section, we perform the tropicalization by using polar coordinates on the fibers. In this case, we will still allow for mixing between the base and fiber coordinates and find that we get a nonzero Hilbert space dimension. Again, we consider $M=T^2$ and its complexification, given by $\CM=T^\star T^2$. We begin with polar coordinates $(r,\theta)$ on the base and fiber coordinates $(R,\Theta)$ on the fiber.  The Maslov dequantization limit \eqref{eqn:Maslov} then takes the form
\begin{align}
    Z_1=e^{r/\hbar+i \Theta}, \quad Z_2=e^{R/\hbar+i\theta}.
\end{align}
We define a new holomorphic 2-form on $\hat{\CM}$
\begin{align}
    \hat{\Omega}=\frac{i}{2\pi}\frac{dZ_1}{Z_1}\wedge \frac{dZ_2}{Z_2}.
\end{align}
In comparison to \eqref{eqn:OldHolo2Form}, we include $i$ to satisfy the conditions mentioned in \secref{sec:BraneQuant} with the new choice of coordinates. Explicitly,
\begin{align}
    \hat{\Omega}=-\frac{1}{2\pi\hbar}\left(dr\wedge d\theta+d\Theta\wedge dR\right)+\frac{i}{2\pi}\left(\frac{1}{\hbar^2}dr\wedge dR+d\theta\wedge d\Theta\right),
\end{align}
which give a different set of real symplectic forms 
\begin{align}
    \hat{\omega}_I&= \frac{1}{2\pi\hbar}\left(dr\wedge d\theta+d\Theta\wedge dR\right)\\
    \hat{\omega}_K&= \frac{1}{2\pi}\left(\frac{1}{\hbar^2}dr\wedge dR+d\theta\wedge d\Theta\right),
\end{align}
with $\hat{\omega}_J$ is then uniquely defined in terms of the other two symplectic forms.
The complex structures also get modified to
\begin{align}
    \hat{I}= \frac{1}{\hbar}\begin{pmatrix}
    0 & 0 & 0 & 1 \\
    0 & 0 & \hbar^2 & 0 \\
    0 & -1 & 0 & 0 \\
    -\hbar^2 & 0 & 0 & 0
    \end{pmatrix}, \quad
    \hat{K}= \begin{pmatrix}
    0 & 0 & -1 & 0 \\
    0 & 0 & 0 & 1 \\
    1 & 0 & 0 & 0 \\
    0 & -1 & 0 & 0
    \end{pmatrix},
\end{align}
with $\hat{J}$ modified similarly as in the previous example. We find that contract to $\mathfrak{iso}(2)$ in this example as well.

In this different setting, $K$ in the $\hbar\rightarrow 0$ limit no longer reduces to a Jordan structure. As a result, the topological A-model is preserved and does not degenerate to a tropological A-model. Additionally, the curvature 2-form $F=\omega_I$ no longer vanishes on the $M$. We find that
\begin{align}
    \text{dim}\,\CH = \int^\oplus_{TT^2}F=\frac{1}{2\pi\hbar}\int d\theta= \frac{1}{\hbar},
\end{align}
where by $\oplus$ , we mean an integral interpreted in the tropical sense along the $r$ direction as explained in  \cite{trsm}.

\section{Conclusions and Future Directions}
\label{sec:Future}

In this work, we extended our previous investigations on tropical branes by constructing explicit examples of tropical coisotropic and tropical Lagrangian branes within the framework of tropological sigma models with boundaries. By carefully analyzing the tropicalized boundary conditions in the sigma model, we demonstrated how these branes emerge naturally from the tropical limit and examined their role in brane quantization. Using the cotangent bundle of the two-torus as an illustrative example, we carried out a brane quantization and computed the corresponding Hilbert space, observing how its dimension deforms under tropicalization.

Notably, we found that certain tropicalization procedures can lead to a vanishing Hilbert space i.e., a zero-dimensional space of quantum states, whereas alternative choices of tropical limits preserve a non-trivial quantization yielding a finite-dimensional Hilbert space. In other words, some methods of taking the tropical limit completely obstruct the quantization, while others allow a well-defined space of quantum states. The behavior of the Hilbert space dimension under tropicalization provides clues that tropical geometry may capture essential features of quantum geometry only in specific limits. 

Although we have derived the presence of these tropical coisotropic and tropical Lagrangian branes through a worldsheet argument, we have not fully investigated how these new objects exist as stable supersymmetric BPS states within the tropicalized target space. The immediate question then becomes how does tropicalization affect supersymmetry? In order to answer this question, we will formulate the tropical analog of spinors and the Dirac equation in the upcoming work \cite{WIP2}. 

The cotangent bundle of the two-torus served as a relatively simple test case. Extending this framework to more general symplectic or hyper-Kähler manifolds, such as higher-dimensional Calabi-Yau manifolds, flag varieties, or moduli spaces of Higgs bundles, is a natural next step. In these more complex phase spaces, one could study tropical coisotropic and Lagrangian branes and determine whether tropicalization preserves key algebraic or geometric structures. This line of inquiry could also shed light on broader connections to mirror symmetry and on the possibility of categorifying brane quantization \cite{Gukov:2010sw, Gukov:2022gei}.

In the standard (non-tropical) topological A-model, Lagrangian and coisotropic A-branes naturally fit into the framework of the derived Fukaya category, and mirror symmetry relates these to objects in the derived category of coherent sheaves \cite{Gukov:2010sw}. A compelling open problem is whether a tropicalized version of these categorical structures can be formulated. For instance, one could ask: Are tropical A-branes objects in a modified or ``tropical'' Fukaya category? How does this category behave under mirror symmetry, and what are the corresponding algebraic objects on the mirror side? These answers have been investigated in the past in the mathematical literature \cite{gross2011tropical} and is consistent with the first pieces of evidence that T-duality is lost in this limit \cite{Albrychiewicz:2024tqe}.

We have presented only a simple example of how tropical brane quantization should work but we have not systematically constructed how the entire framework should appear. In conventional brane quantization, one generically focuses on affine varieties that are defined by a finite set of polynomials for a finite set of complex variables. One might wonder if tropical brane quantization is the correct quantization procedure to use when the affine varieties are instead reduced down to tropical varieties. From there, investigating how various choices of tropical branes affect the quantization process could reveal new insights into the uniqueness of tropical brane quantization.  

\acknowledgments
We wish to thank Ori Ganor and Petr Ho\v{r}ava for useful discussions. This work has been supported by the Berkeley Center for Theoretical Physics.

\nocite{*}

\bibliographystyle{JHEP}
\bibliography{ticfpiqm.bib}

\end{document}